\documentclass[]{spie} 

\usepackage{cite}
\usepackage[]{graphicx}
\usepackage{amssymb} 
\title{ZnS/Diamond Composite Coatings for Infrared Transmission Applications Formed by the Aerosol Deposition Method}
\author{Scooter D. Johnson\supit{a}, Fritz J. Kub, and Charles R. Eddy, Jr.
\skiplinehalf
Electronics Science and Technology Division, U.S. Naval Research Laboratory, Washington, D.C., 20375, U.S.A.}
\authorinfo{Further author information: (\supit{a}American Society of Engineering Education Postdoctoral Fellow residing.  Send correspondence to S.D.J.: E-mail:scooter.johnson.ctr@nrl.navy.mil, Telephone: 1 202 404 5622)}
\begin{document}
\maketitle
\begin{abstract}
The deposition of nano-crystalline ZnS/diamond composite protective coatings on silicon, sapphire, and ZnS substrates, as a preliminary step to coating infrared transparent ZnS substrates from powder mixtures by the aerosol deposition method is presented. 
Advantages of the aerosol deposition method include the ability to form dense, nanocrystalline films up to hundreds of microns thick at room temperature and at a high deposition rate on a variety of substrates. 
Deposition is achieved by creating a pressure gradient that accelerates micrometer-scale particles in an aerosol to high velocity.
 Upon impact with the target substrate the particles fracture and embed. 
Continued deposition forms the thick compacted film. 
Deposition from an aerosolized mixture of ZnS and diamond powders onto all targets results in linear trend from apparent sputter erosion of the substrate at 100\% diamond to formation of a film with increasing fractions of ZnS. 
The crossover from abrasion to film formation on sapphire occurs above about 50\% ZnS and a mixture of 90\% ZnS and 10\% diamond forms a well-adhered film of about 0.7 $\mu$m thickness at a rate of 0.14 $\mu$m/min.  
Resulting films are characterized by scanning electron microscopy, profilometry, infrared transmission spectroscopy, and x-ray photoemission spectroscopy. These initial films mark progress toward the future goal of coating ZnS substrates for abrasion resistance.  
\end{abstract}
\keywords{Aerosol Deposition Method, Zinc Sulfide, Diamond, Abrasion Resistance, Protective Coating, Thick Films, Infrared Transmission}
\section{Introduction}
Many infrared (IR) sensor applications require optics with good tranparency in the 3 -- 5 and 8 -- 14 $\mu$m wavelength range and good mechanical performance against harsh environmental conditions such as rain and sand impact and thermal shock\cite{harris-materials}.
Of the known materials the so-called ``two-color"  or ``multispectral"  material ZnS is a good candidate due to strong transmittance in this desired range  even at high operating temperature due to its large bandgap of 3.6 eV.
Unfortunately, its poor mechanical properties require a durable protective coating.
Diamond is a strong candidate as a protective coating material for several reasons.
It has good transparency in the long-wave IR and has mechanical properties orders of magnitude better than other optical materials.
It has twice the mechanical strength of ZnS and over 70 times its thermal conductivity.
A measure of thermal shock at mild heating, the Hasselman figure of merit, for diamond is 200 times better than ZnS.
For these reasons there has been much focus on depositing diamond and diamond-like protective coatings onto ZnS, but processes such as chemical vapor deposition require high temperature or other operating environments inimical to the ZnS structure.
While there have been some advancements in coating ZnS with a hard diamond layer by plasma laser discharge\cite{davanloo-protective,davanloo-adhesion} and diamond composites by filtered cathode vacuum arc\cite{zhu-multilayer} there is to date no current technique that meets the current application needs.
An improved deposition technique would deposit diamond and/or diamond-containing composites at low temperature, build up a several micron-thick dense film that is well adhered to the substrate, offer improved abrasion resistance, and cover a large surface area.
For this reason the novel aerosol deposition method (ADM) has been employed to potentially solve this long-standing problem.

The ADM is a film fabrication technique that utilizes an impact solidification phenomenon of fine particles that was invented in 1997  by Dr. Jun Akedo.
The ADM has demonstrated the most success of any technique for depositing ceramic and metallic compounds \cite{akedo-aerosol,akedo-firing,imanaka-aerosol,sahner-assessment} such as, $\alpha$-Al$_2$O$_3$, (Ni,Zn)Fe$_2$O$_3$, AlN, MgB$_2$, BaTiO$_3$, SrTi$_{0.7}$Fe$_{0.3}$O$_{3-\delta}$ onto a substrate, such as Cu, Al, glass, stainless steel, Si, SiO$_2$ to form thick  ($>$ 100 $\mu$m) dense (95\% bulk density) films at room temperature.
Recently, ADM has been used to deposit mixtures of Al$_2$O$_3$ and nano-diamond onto glass for anti-scratch and smudge resistance\cite{lee-al2o3}.

\begin{figure}[tb]
\begin{center}
\scalebox{.5}{\includegraphics{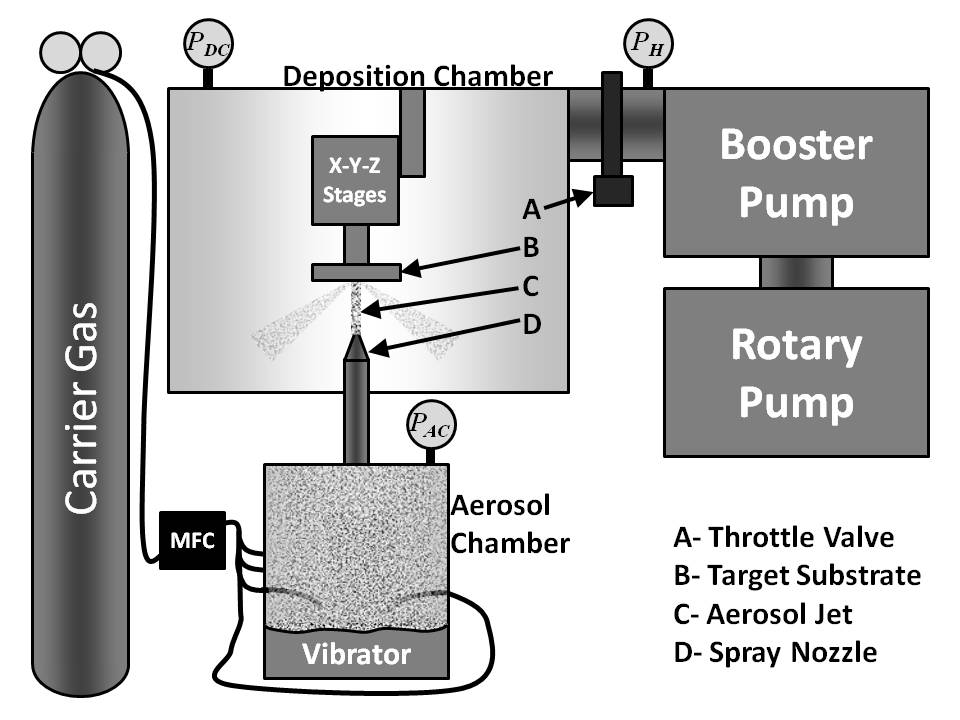}}
\caption{Main components in the NRL ADM system: carrier gas source, mass flow controller (MFC), aerosol chamber, powder vibrator, deposition chamber, X-Y-Z translation stages, and pumps. Pressure monitored at the locations marked $P_{AC}$, $P_{DC}$, and $P_{H}$ for the aerosol chamber, deposition chamber, and pump head, respectively. }
\label{admsys}
\end{center}
\end{figure}

\begin{figure}[tb]
\begin{center}
\scalebox{.5}{\includegraphics{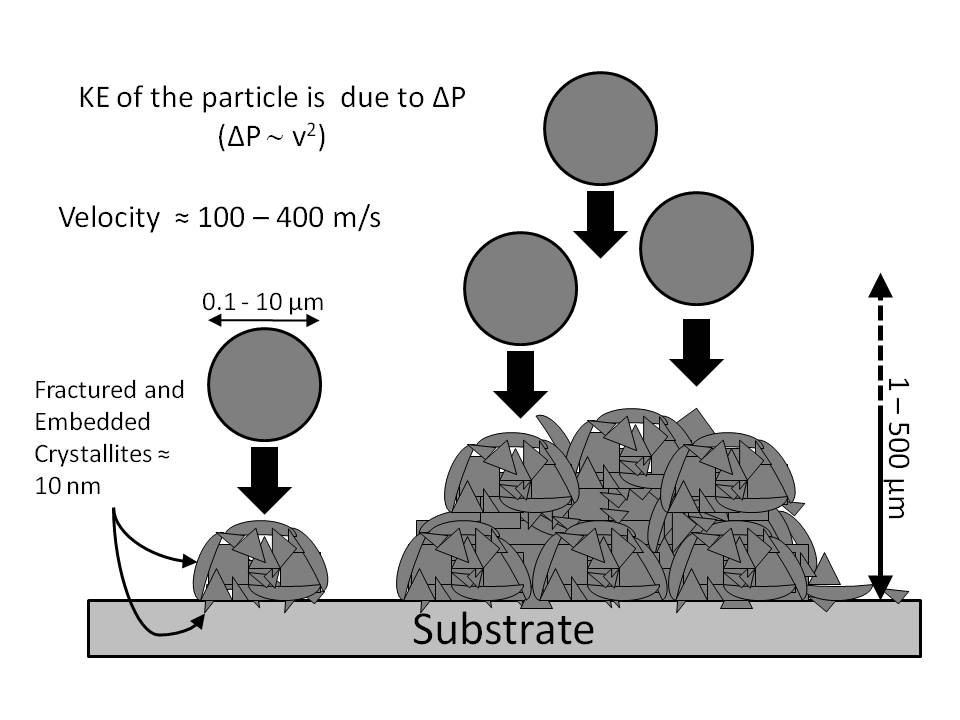}}
\caption{Illustration of the deposition process.  Initial particles that impact the substrate fracture and embed forming the anchoring layer.  Subsequent particles impact, fracture, and compact the film resulting in a thick dense film of crystalline nanoparticles 10 - 100 nm in size.}
\label{dep}
\end{center}
\end{figure}

There are four main components of the ADM system: 1) an aerosol chamber (AC); 2) a deposition chamber (DC); 3) carrier gas source; and 4) a vacuum pump as shown in Figure \ref{admsys}.
In this process a chamber containing an aerosol of fine (0.1 -- 10 $\mu$m) particles is pressurized with nitrogen, helium, or oxygen while a chamber containing a target substrate is pumped to $<$ 1 Torr.
The pressure differential $\Delta$P ($\sim$ kinetic energy) accelerates the aerosolized particles through a nozzle toward the substrate.
Previous reports of empirically measured velocities are 150 -- 550 m/s for Pb(Zr$_{0.52}$,Ti$_{0.48}$)O$_3$  (PZT)\cite{akedo-room} and 650 m/s for ultrafine silver particles\cite{lebedev}.
Figure \ref{dep} illustrates the currently accepted deposition mechanism.
The film begins to form when incident particles impact, fracture, and embed into the substrate.
This impact  causes indentation and abrasion of the target area giving rise to an increased surface area which facilitates formation of a well-adhered anchor layer comprised of 10 -- 100 nm sized particles.
Subsequent impacts serve to compact the underlying film and bonds the crystallites.
Since only fracturing occurs, the fabricated film has the same crystalline structure as the raw powder.
The structure of the film is a comprised of densely packed crystalline nanoparticles held together by what is thought to be close-range mechanical and chemical interactions mediated by fracture, and/or plastic deformation of the particles\cite{akedo-room}.
These films are more than 95\% of the density of the bulk material\cite{akedo-aerosol}.
Several reports claim that the deposited particles retain the crystalline microstructure, although signs of strain in the deposited particles can also occur\cite{sahner-assessment,imanaka-aerosol,nam-alumina,akedo-room,kato-magnetic}.
Finite element simulations suggest that the pressure and temperature at the impact site is not sufficient to cause melting \cite{akedo-room,akedo-jet}, but to date a more thorough investigation of the mechanisms of adhesion in this process has not been reported.

The ability of the ADM to form dense thick films at room temperture over a broad range of material systems makes this novel method a potential solution for coating diamond or diamond-containing composite films onto ZnS substrates.
We present the data for deposition onto ZnS, silicon, and sapphire substrates.
While the immediate motivation is to coat ZnS, by comparing the deposition results of the three system a broader understanding of the deposition mechanism may be ascertained.

\section{Materials \& Methods}
The ADM used in this work is a  custom built system at the U.S. Naval Research Laboratory, Washington, D.C. (NRL).
Typically 100 -- 200 g of powder is  loaded into the AC, a modified virtual cyclone concentrator powered by a  PITT-3 aerosol generator ({\em Alburty Labs}, Drexel, MO 64742).
A programmable Hewlett Packard 8112A 50 MHz function generator sweeps frequency  between 130 -- 150 Hz to keep the powder agitated.
A 1/2 inch diameter tube connects the AC to the DC, where the tube is capped with a spray nozzle ({\em Comco, Inc.}, Burbank, CA 91504) of rectangular aperture 0.016 by 0.190 inches (0.41 $\times$ 4.83 mm).
The target substrate is mounted with double-sided copper tape to an aluminum base which is in turn mounted to a system of computer-controlled stages ({\em Physik Instrumente, GmbH \& Co. KG}, D-76228 Karlsruhe/Palmbach) that provide {\em X, Y}, and {\em Z} coordinate translation.
An Edwards E1M 275 rotary pump and EH 1200 booster pump provide the vacuum required to create the pressure difference $\Delta P=P_{AC}- P_{DC}$ that drives the aerosolized powder.
The base pressure of the system is 3 mTorr.
For a 5 minute deposition the total run time was typically about 30 minutes.
The design of the system was modeled using VacTran modeling software ({\em Professional Engineering Computations}, Livermore, CA) for optimal operation at flow and pressure conditions reported by other ADM researchers in the literature\cite{akedo-room,lee-al2o3,kato-magnetic,sahner-assessment}.
Table \ref{table-compare} compares  initial flow tests of the NRL ADM to typical values of reported systems. 
\begin{table}
\caption{Comparison of reported operating parameters between established ADM and NRL ADM systems. }
\begin{center}
    \begin{tabular}{l c c}
        \hline
        Parameter & Reported\cite{akedo-room} ADM & \hspace{15pt} NRL ADM \\ \hline
	Pres. in DC (Torr)            &0.4\textendash 15                                            &0.2\textendash1\\ 
	Pres. in AC (Torr)            &75\textendash600                                           &50\textendash800\\
	Nozzle Aperture (mm$^2$)   & 5$\times$0.3, 0.4$\times$10       & 5$\times$0.4\\
	Carrier Gas                     &He, N$_2$, air                                  &N$_2$\\
	Gas Flow Rate (l/min.)    & 1\textendash10                                               & 1\textendash20 \\
	Dep. Area (mm$^2$)              & 40$\times$40, 400$\times$400  & 25$\times$50 \\
	Sweep Speed (mm/s)      & 0.125\textendash10                                       & 0.05\textendash0.65 \\
	Nozzle-Target Dist. (mm)& 1\textendash40                                               & 0.5\textendash50 \\
        \hline
    \end{tabular}
\label{table-compare}
\end{center}
\end{table}

Deposition targets were p-type, boron-doped $<$100$>$  single-side polished silicon wafers 380 $\mu$m in thickness,
 a-plane double-side polished sapphire wafers 320 $\mu$m in thickness,
and ground ZnS wafers 508 $\mu$m in thickness)
.
The wafers were diced to 1.5 $\times$ 1.5 cm squares and masked with cellophane tape.  The deposition target area was exposed by cutting out a 2 $\times$ 10 mm section using a stencil and razorblade (see inset of Figure \ref{80ftir}).
Synthetic grade metal bond diamond powder 1 -- 6 $\mu$m in size ({\em DIDCO Inc.}, New York, NY 10010) and ZnS powder $<$ 5 $\mu$m in size 99.99 \% pure ({\em Noah Technologies Corp.}, San Antonio, TX 78249) were used in this work.
 Weight percent refers to percentage of ZnS in the mixture. 
Weight was measured on a Sartorius ED224S balance with a 220 g capacity and resolution of 0.1 mg.
 The mass of the starting powder in the AC was between 94 -- 191 g.
  A plot of film thickness versus starting powder mass shows no correlation between the powder quantity and the resultant films characteristics.
Deposition target area step height was  measured on a Tencor profilometer with a stylus force of 7.9 mg, scan length 3 mm at 0.1 mm/s, and 100 Hz sampling rate.
The horizontal resolution was 1 $\mu$m and the vertical range/resolution was either 13 $\mu$m/1 \AA  \hspace{5 pt}or 300 $\mu$m/25 \AA.
A typical sample was profiled at three locations (near top, middle, and bottom) across the width of the target area.
 For each profile the root mean square (rms) roughness, maximum height, minimum height, and average step height was tabulated.
The  latter being determined by comparing the average height of the deposited region to the height of the undeposited region in the scan.
We determined an uncertainty of 100 \AA \hspace{5 pt}in roughness and step height experimentally.
Scanning electron micrographs (SEM) were taken using a Leo Supra 55 scanning electron microscope with maximum accelerating voltage of 30 kV.
Fourier Transform Infrared (FTIR) transmission measurements were performed on a Thermo Scientific Nicolet Continu$\mu$m IR microscope with OMNIC acquisition software using a KBr beam splitter and an aperature of 35 $\mu$m using a narrow band 250 $\mu$m MCT detector.
   The spot size for each transmission measurement was 150  $\times$ 150 $\mu$m. 
  64 scans were taken at each point with a resolution of 8 $\mu$m.  An {\em auto atmospheric suppression} algorithm was used to correct for CO$_2$ and H$_2$O absorption. 
The index of refraction of the system was set to sapphire (n = 1.77) with a thickness of 320 $\mu$m. 
 The sapphire substrate background was measured for each sample and subtracted to obtain transmittance from the target area only.
Background signals of each sample substrate had nearly identical characteristics.
Each target area was measured at 4 or 5 distinct locations (see inset of Figure \ref{80ftir}).
 X-ray photoemission Spectroscopy (XPS) was performed on a Thermo Scientific K-Alpha system controlled by Avantage software. 
Each Element was scanned 20 times in 0.15 eV steps. 
XPS data was background subtracted using the Tougaard algorithm and peaks fit with Unifit software.

The settings used for the ADM system, unless otherwise noted: 
$\Delta$P = 200 Torr, 
target area = 2 $\times$ 10 mm (masked),
carrier gas =  nitrogen (99.999\% minimum purity),
deposition time = 5 minutes,
scan speed = 0.5 mm/s,
nozzle-substrate distance = 7.5 mm.

\section{Results}
\subsection{Surface Morphology}
\begin{figure}[tb]
\begin{center}
\scalebox{.33}{\includegraphics{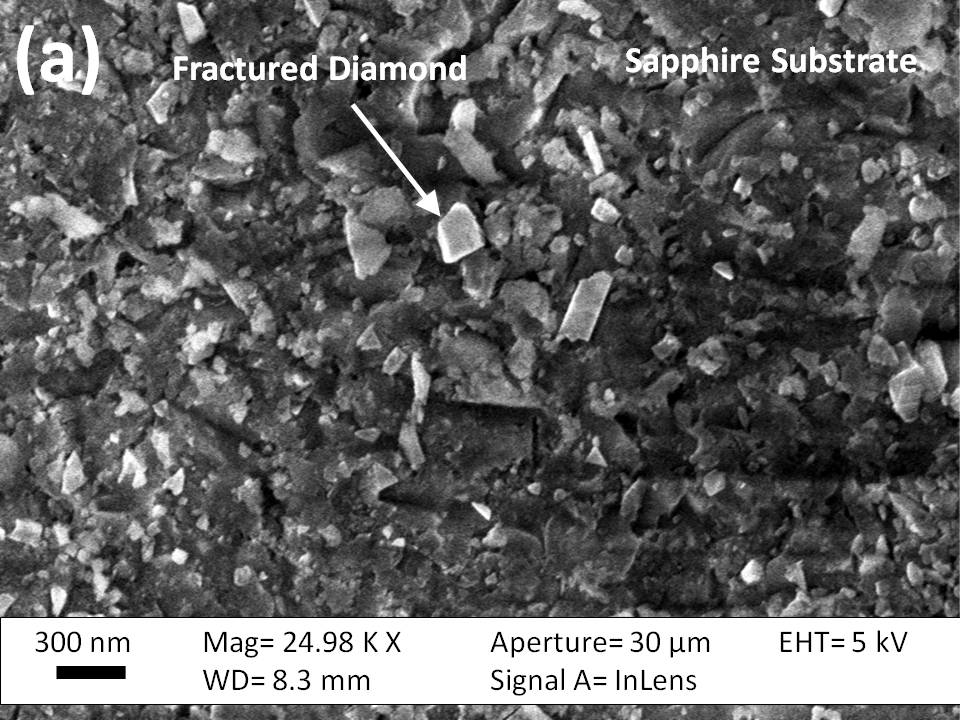}\includegraphics{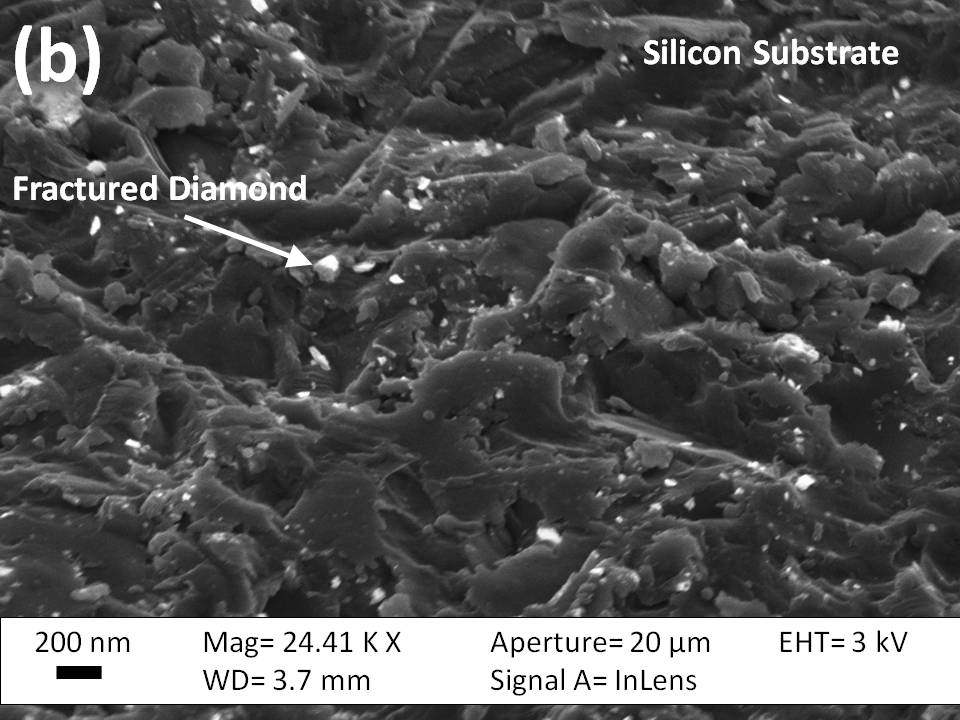}} \\
\scalebox{.33}{\includegraphics{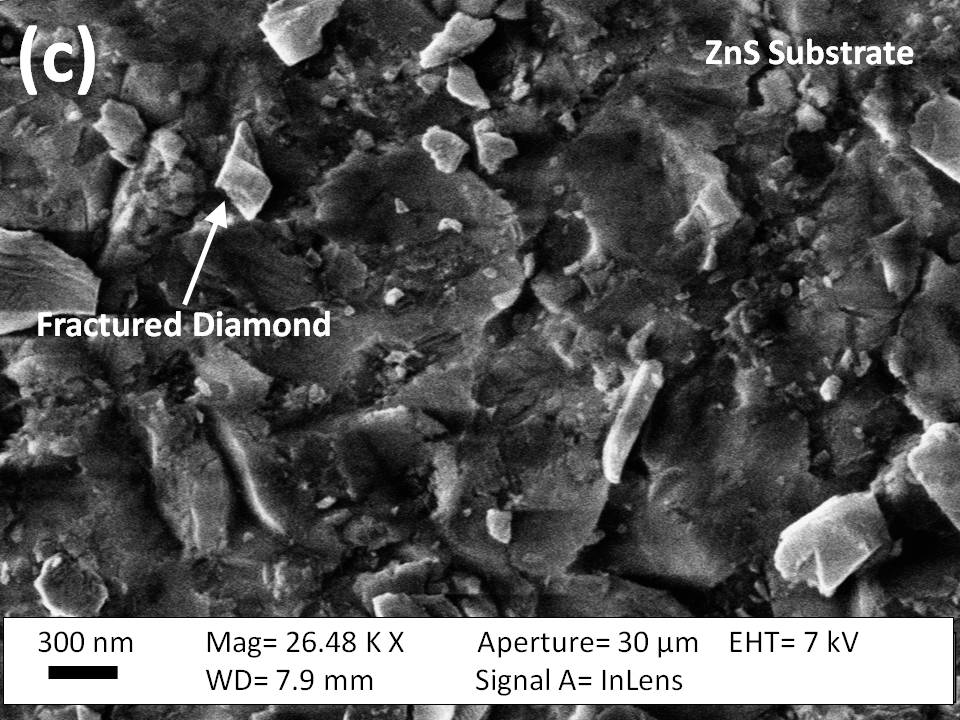}\includegraphics{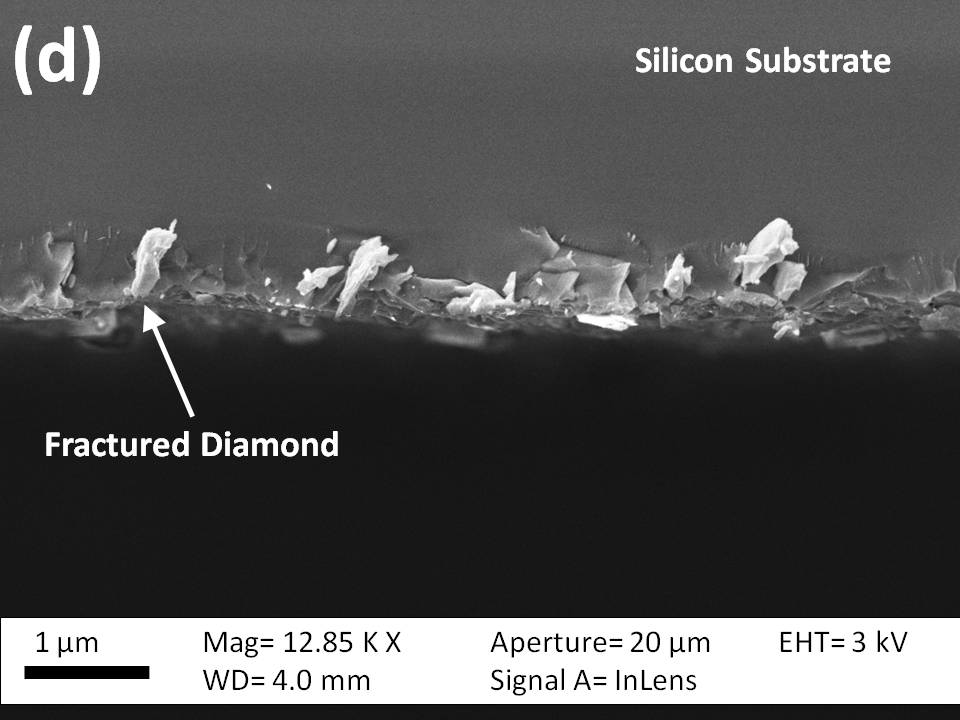}}
\caption{SEM images of the target area after deposition using only diamond particles. (a) Sapphire substrate. (b) Silicon substrate tilted at a 45-degree angle to highlight the damage to the substrate. (c) ZnS substrate. (d) Cross-section of fractured diamond embedded about 0.5 $\mu$m into silicon substrate from (b).}
\label{dmd-dep}
\end{center}
\end{figure}

Figure \ref{dmd-dep} (a), (b), and (c) are images of the surface of the target area after pure diamond deposition onto sapphire, silicon, and ZnS, respectively. 
The silicon surface in (b) is tilted at a 45-degree angle to highlight the damage.   Several features are evident from these images: 1) The diamond particles are fractured to at least 1/3 the initial size of the smallest particles used.  
Many of the visible fragments appear much smaller still. 2)  The diamond particles do not form a continuous film. 3) The substrate surface is severely damaged. 
The substrate surface, which is peeled and cracked shows signs of plastic deformation\cite{pharr-new,callahan-extent,kollenberg-plastic}. 
In (d) a cross-section of the silicon sample in (b) further highlights the damage of diamond particle impact extending about 0.5 $\mu$m into the substrate.
The extent of the surface damage evident in these figures is confirmed by profilometry (see Figure \ref{thick} and accompanying text). 
These images are representative of similar results when using longer deposition times, lower or higher $\Delta$P (kinetic energy), and diamond particles of sizes from $<$ 0.25 to 10 $\mu$m in the NRL ADM system.  
These initial depositions motivate introducing a ``binder'' material that could facilitate adhesion to the substrate and help absorb the initial impact of the diamond.  In addition, a suitable material with a better thermal match to the ZnS substrate might decrease film delamination from thermal cycling.  An obvious material to choose is ZnS powder.

\begin{figure}[tb]
\begin{center}
\scalebox{.33}{\includegraphics{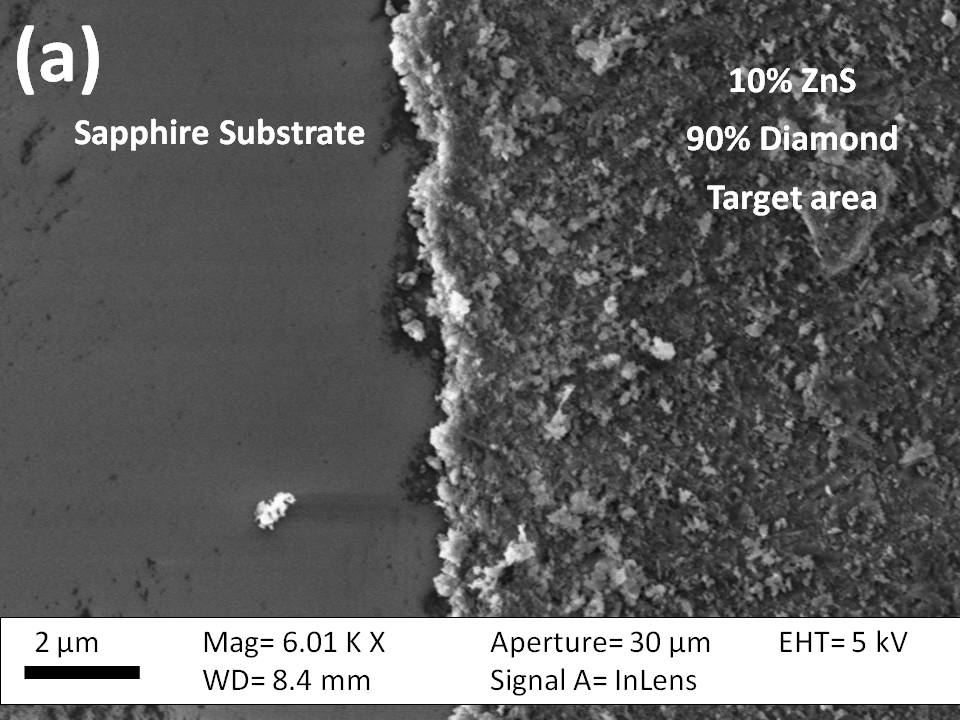}\includegraphics{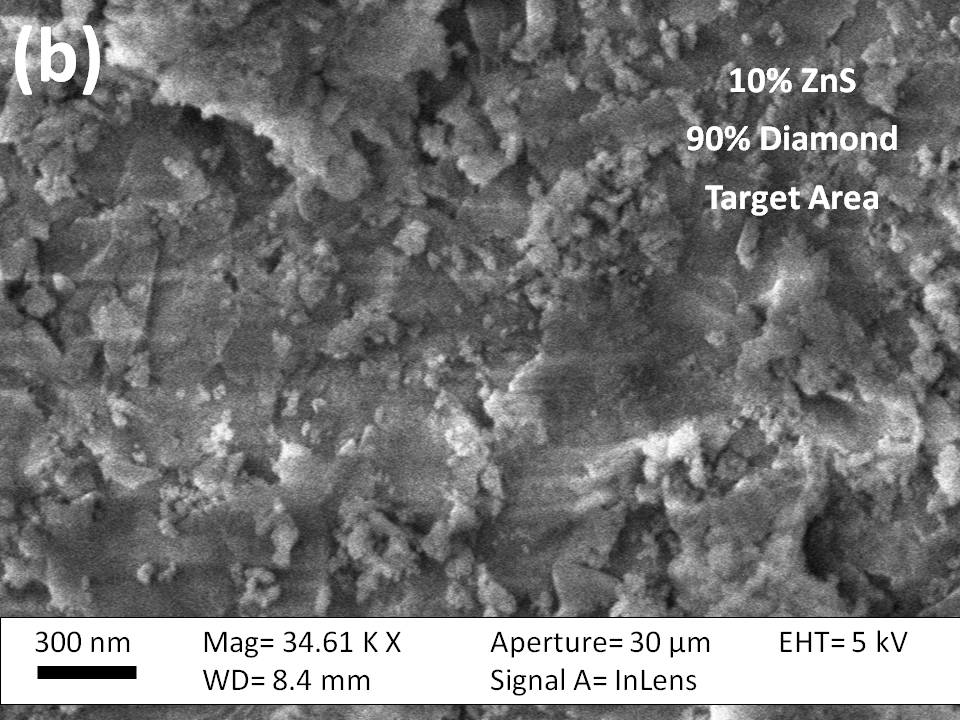}}\\ \vspace{-2 pt}
\scalebox{.33}{\includegraphics{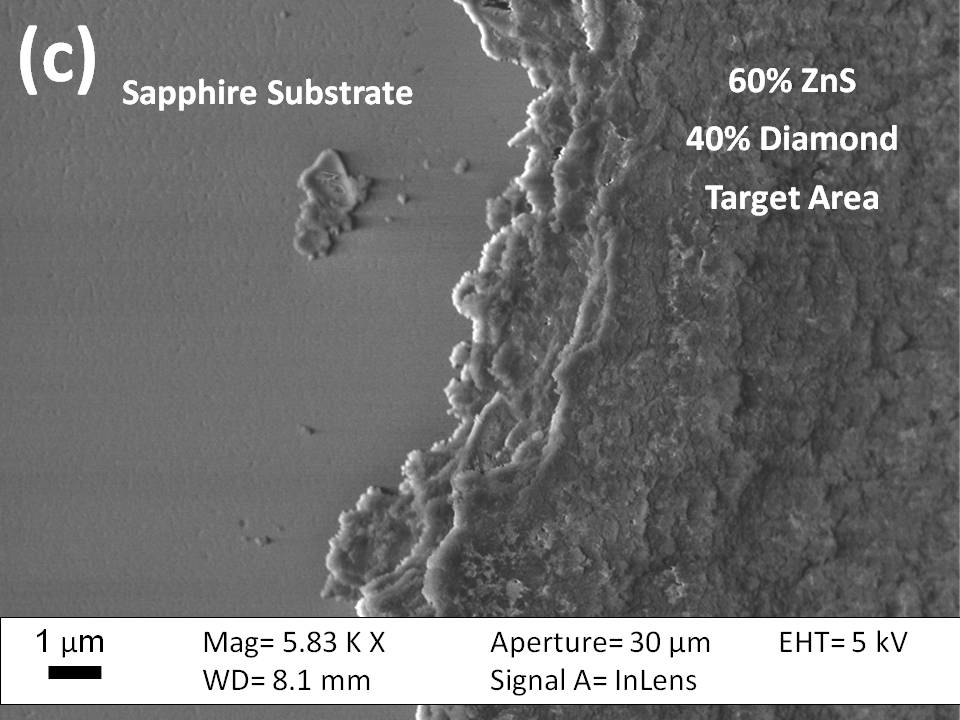}\includegraphics{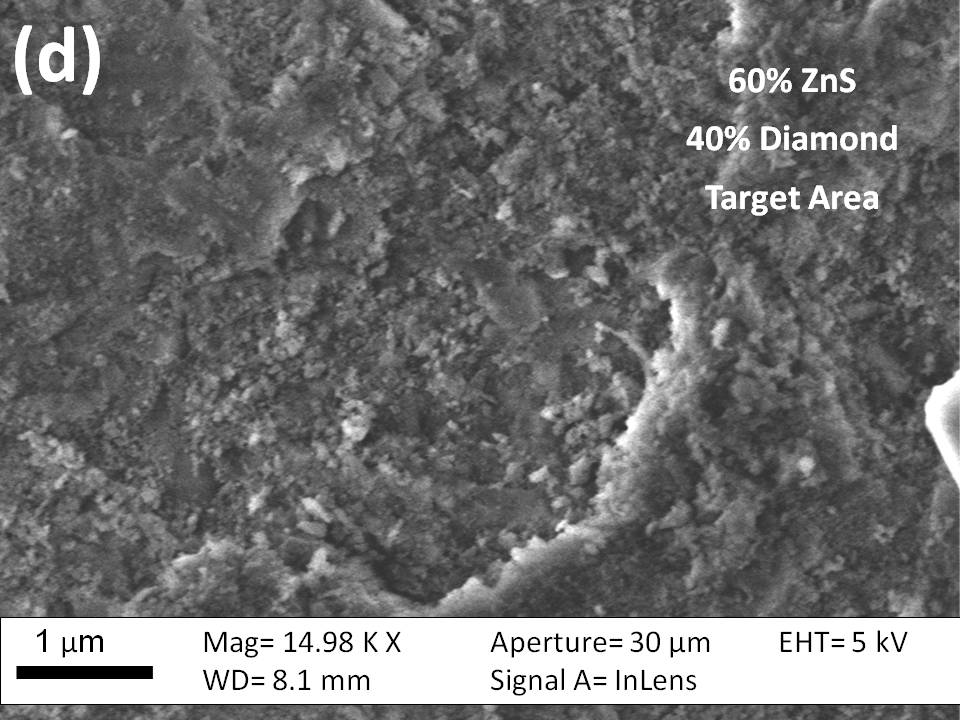}}
\caption{SEM images of  post-deposition surface after using 10\% and 60\% ZnS mixtures on sapphire. (a) Edge of the target area after using a 10\% mixture. (b) Magnified image of the target area after using a 10\% mixture. (c) Edge of the target area after using a  60\% mixture. (d) Magnified image of the target area after using a 60\% mixture.}
\label{mix-dep}
\end{center}
\end{figure}

\begin{figure}[tb]
\begin{center}
\scalebox{.5}{\includegraphics{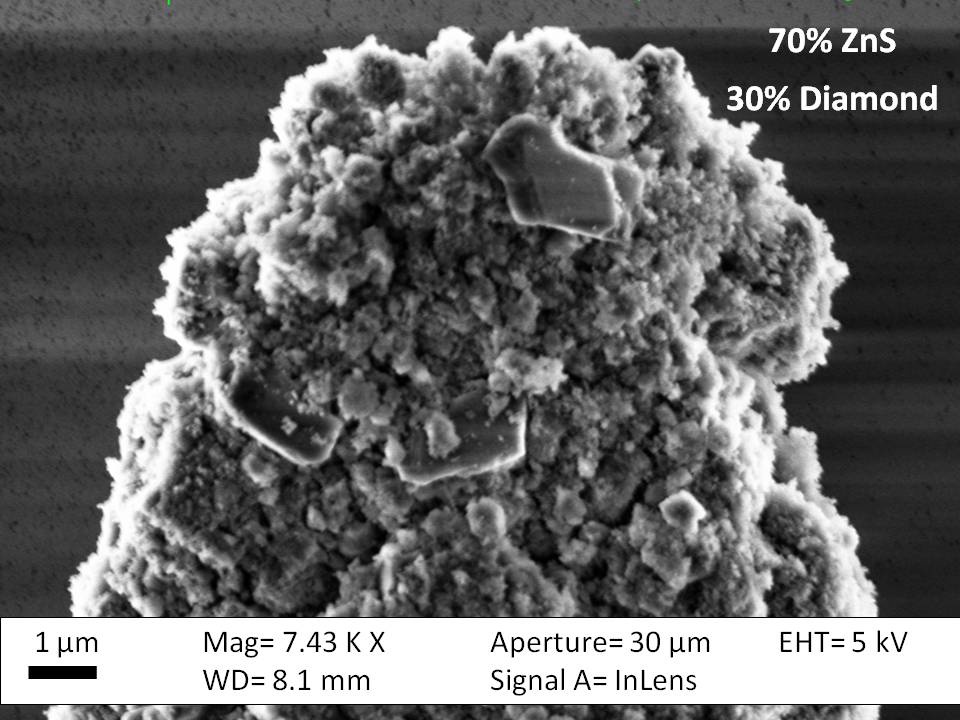}}
\caption{SEM image of a fragment of film formed from a 70\% mixture showing $\sim$ 1 $\mu$m sized diamond conglomerated with ZnS.}
\label{conglom}
\end{center}
\end{figure}

Figure \ref{mix-dep} shows SEM images of deposition onto sapphire using 10\% and 60\% mixtures. 
The images in (a) and (c) show the edge of the target area in contrast to the bare sapphire outside of the target area.
 The granularity of the film in (a) is presumably due to the large amount of diamond content. 
 In (c) the film surface appears less granular presumably due to the increased ZnS content. 
 In (b) and (d) a magnified view of the films show roughness consistent with observed substrate damage for depostion with diamond only.  
At higher magnification the granularity seen in (a)  is not apparent in (b) and the film surface morphology between the varying mixtures becomes essentially indistinguishable.  
Figure \ref{conglom} shows a delaminated portion of a film deposited with a 70\% mixture that has broken away from the edge. 
 A well-compacted conglomerate of larger diamond shards and the indistinguishable ZnS particles illustrate that the film is well compacted and comprised of both ZnS and diamond even at 70\% ZnS content.
\subsection{Surface Profile}
\begin{figure}[tb]
\begin{center}
\scalebox{.45}{\includegraphics{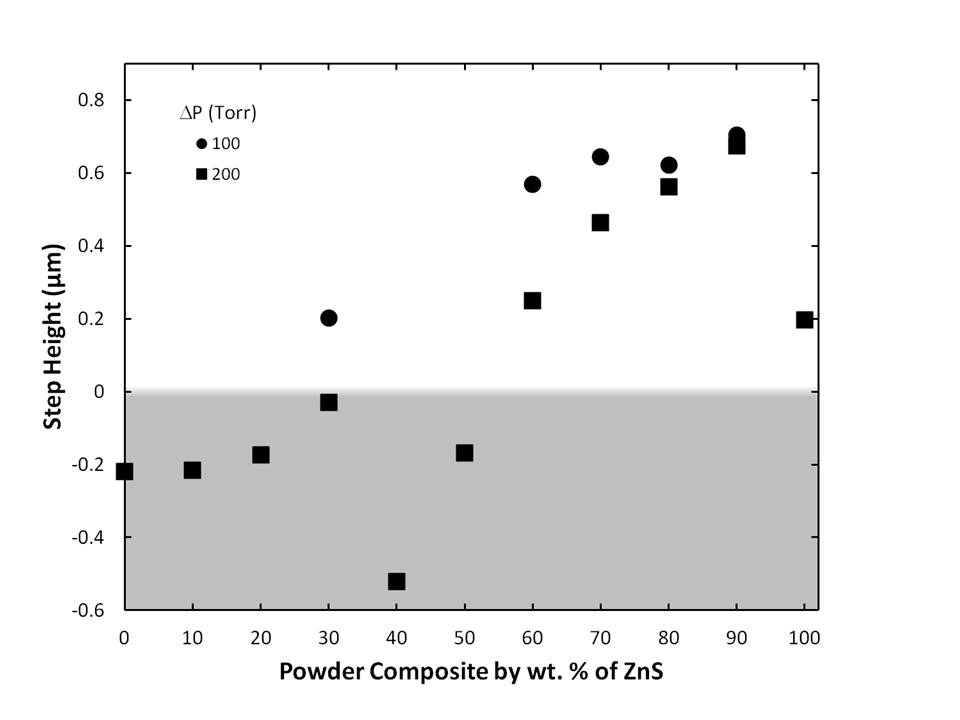}}
\scalebox{.45}{\includegraphics{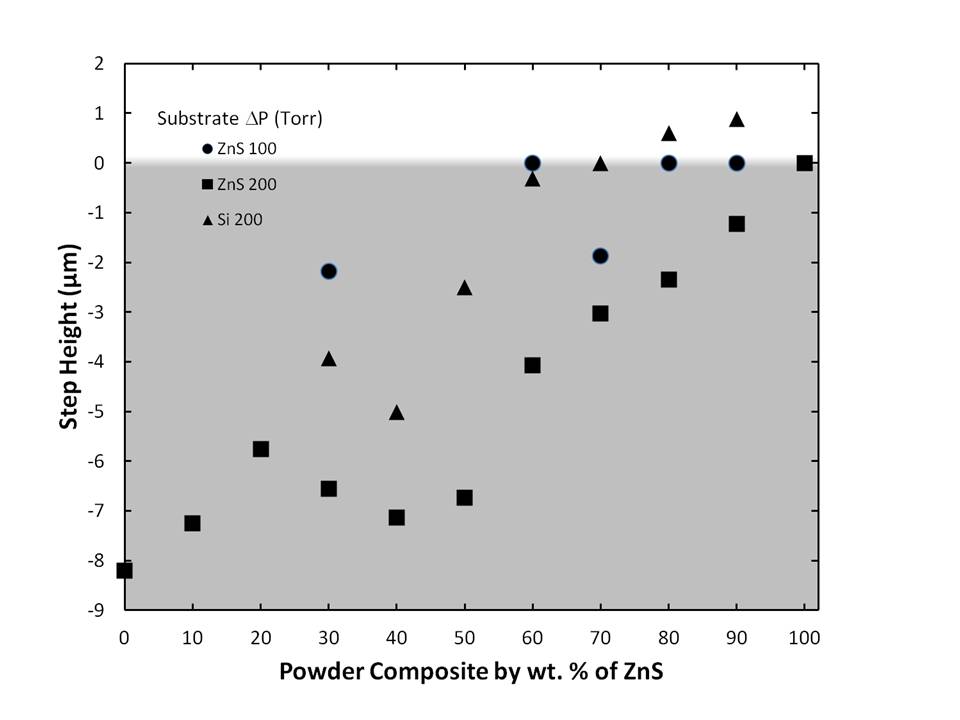}}
\caption{Plots of average step height taken across three locations of the target area versus powder mixture percentage. The grey region highlights where a negative step height was measured. The legend is pressure difference ($\Delta$P) . (Top) Step height across target area on sapphire at two values of $\Delta$P. (Bottom) Step height across target area of ZnS and silicon.}
\label{thick}
\end{center}
\end{figure}

Figure \ref{thick} are plots relating the film step height to the powder mixture percentage. 
Each point represents the average of three profiles measured at distinct locations along the width of the film.  
Negative step heights are emphasized by the shaded region.
The top plot, is step height measured after deposition onto sapphire for two values of $\Delta$P. 
For $\Delta$P = 200 Torr the step height becomes positive for mixtures from 60\% to 90\% increasing steadily at about 14 nm per wt. \%.  
Overall, the film thickness increases at a rate of about 11 nm per wt. \%. 
Below 60\% all step heights are negative values indicating erosion of the substrate surface. 
For $\Delta$P = 100 Torr the step height remains positive for all mixtures and increases at 8 nm per wt. \% coinciding with the $\Delta$P data above 70\%.
The plot on the bottom is step height measured after deposition onto ZnS and silicon.
The deposition on silicon was done at $\Delta$P = 200 Torr with a negative step height below 70 \% mixture.
Although the maximum step height is 0.9 $\mu$m at 90\% the formation of a positive step height does not occur until 70\% mixture is used and the maximum negative step height of about -4 $\mu$m at 30\% mixture is much larger than on sapphire.
For deposition using $\Delta$P =  100 and 200 Torr onto ZnS substrates only negative step height values are observed.  
\subsection{Transmittance}
\begin{figure}[tb]
\begin{center}
\scalebox{.5}{\includegraphics{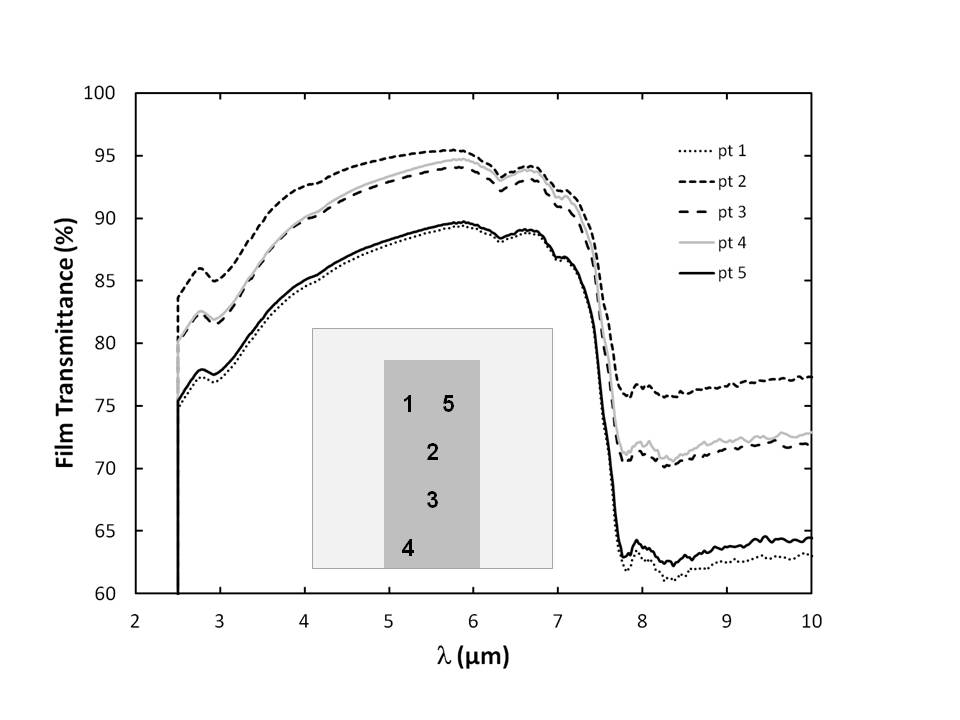}}
\caption{Film transmittance of 80\% mixture on sapphire with $\Delta$P = 200 Torr at 5 locations on the target deposition area.  The inset shows a drawing of the sapphire substrate and the strip resulting from deposition at the target area in grey.  The numbers correspond to the locations of the transmittance data in the plot.}
\label{80ftir}
\end{center}
\end{figure}

Figure \ref{80ftir} is a plot of transmittance between 2 -- 10 $\mu$m for 5 locations in the target area after a deposition onto sapphire with an 80\% mixture at $\Delta$P = 200 Torr. 
The inset drawing is of the sapphire substrate (light grey) and the target area (grey) with the locations of the measured transmittance indicated.
The sharp drop in transmittance above 7.8 $\mu$m and below 2.5 $\mu$m  is due to the strong absorption of sapphire. 
As indicated by the transmittance at points 1 and 5, film thickness or roughness may be non-uniformily deposited across the target area resulting in the observed transmittance profile.

\begin{figure}[tb]
\begin{center}
\scalebox{.45}{\includegraphics{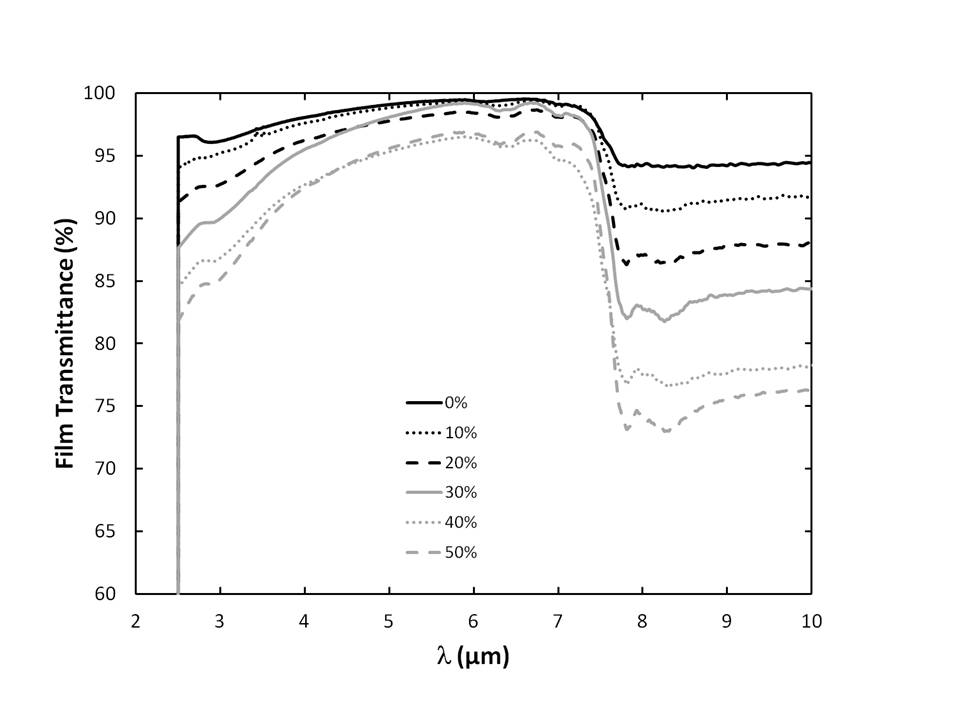}}\vspace{-15 pt}
\scalebox{.45}{\includegraphics{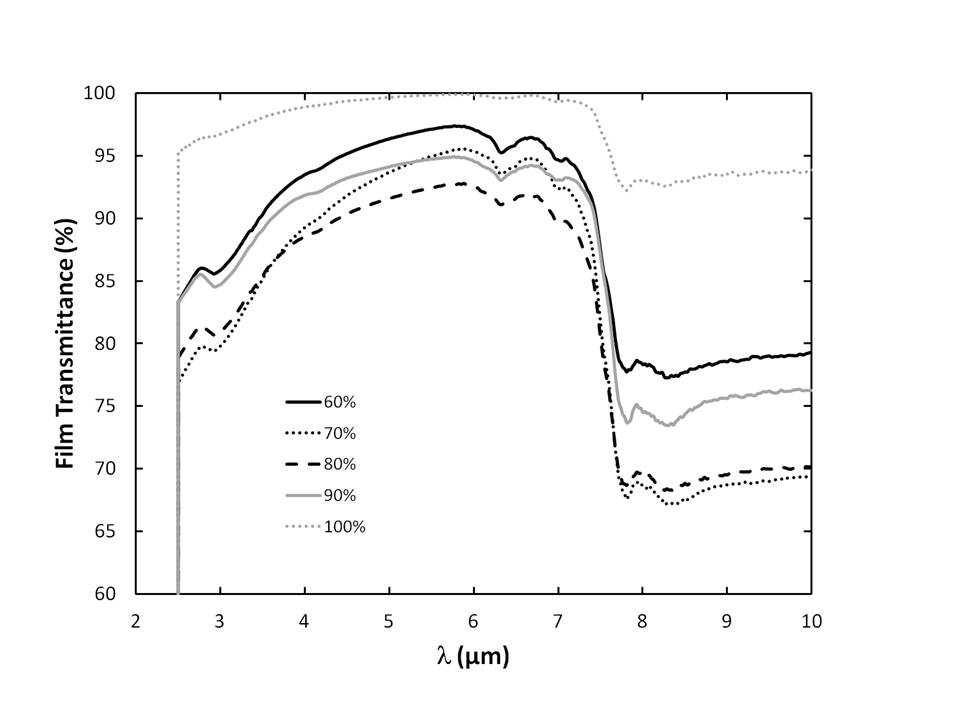}}
\caption{Transmittance of target area for different compositions. (Top) Transmission data for target areas with negative step height. (Bottom) Transmission data for target areas with positive step height.}
\label{ftir}
\end{center}
\end{figure}

Figure \ref{ftir} shows plots of averaged transmittance curves for each mixture. 
The plot on the top (bottom) shows transmittance of the target area with a negative (positive) step height.  
The data in the top plot show a general trend toward decreasing transmittance with increasing percent of ZnS powder and (less negative) step height.  
By comparison, the data on the bottom show a general trend toward decreasing transmittance with increasing step height only up to 80\%. 
 At 90\% the transmittance increases, suggesting a trend not only with step height, but with composition.

\subsection{X-ray Photoemission Spectroscopy}
\begin{figure}[tb]
\begin{center}
\scalebox{.5}{\includegraphics{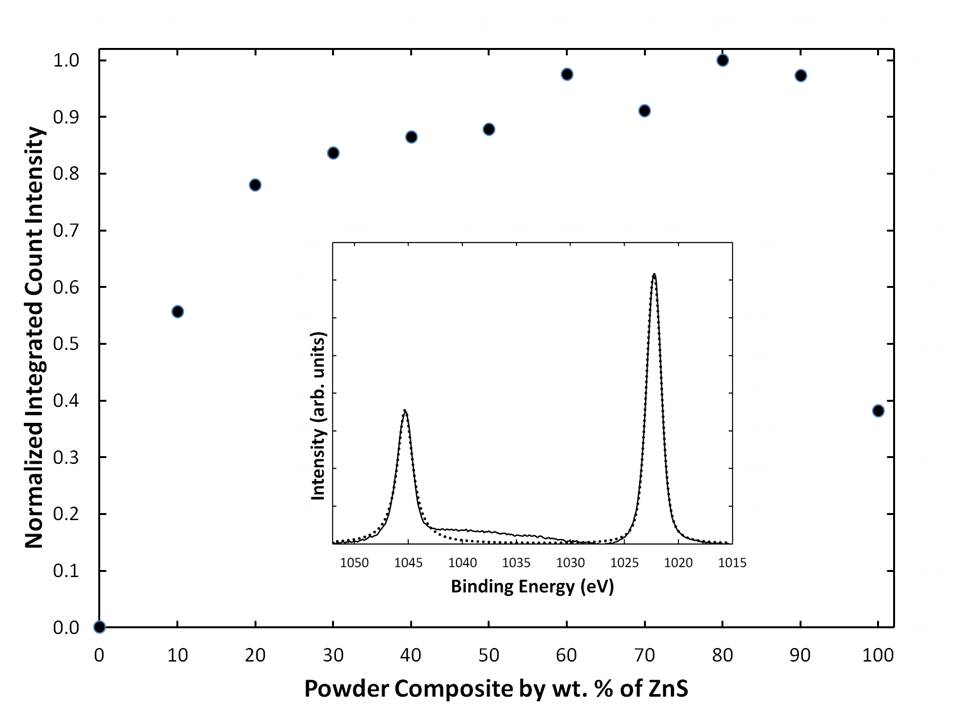}}
\caption{Normalized integrated count intensity of the Zn $2p$ doublet for each mixture.  
The inset is an intensity plot of a Zn $2p$ doublet in the target area of a 30\% mixture with background subtracted and is representative of data and fits for other mixtures.
 Data is shown as a solid line.  The dotted line is the fit to the doublet.}
\label{xps}
\end{center}
\end{figure}

Figure \ref{xps} is a plot of the normalized integrated count intensity of the Zn $2p_{1/2}$ and $2p_{3/2}$ peaks in the target area for each mixture as deposited onto sapphire. 
The inset is a representative signal showing the Zn $2p$ doublet with background subtracted and dotted fit line overlayed.
 As the main plot indicates, the total count increases with increased Zn percentage as one might expect. Note the 0\% mixture at the origin.
  The sudden drop at 100\% could be due to poor coverage due to poor adhesion.
 At percentages below 60\% where a negative step height might suggest a purely abrasive mechanism there remains a fairly strong Zn signal. 
 The presence of Zn, ergo ZnS, indicates that other mechanisms might be at work than pure abrasion of the substrate.

\section{Discussion and Conclusion}
Deposition with pure diamond produces a substrate that is abraded, surface roughened, and sparsely embedded with the particles. 
While diamond-on-diamond adhesion in this process would be desirable, the interaction may be very elastic.  
The bounce kinetic energy $KE_b \propto (1-e^2)/e^2$ is the kinetic energy required for a bounce to occur between two interacting objects\cite{hinds-aerosol,dahneke-capture}.
  Here $e$ is the coefficient of restitution as determined by the ratio of the difference in the incoming and outgoing velocities. 
 For an impact between two nearly perfectly elastic objects such as diamond particles $e \sim 1$ resulting in diamond bounce and/or fracture, but not adhesion. 
 Serendipitously, the impact and damage resulting from the diamond may actually enhance deposition by softer materials, such as ZnS, by increasing the surface area and forming regions where particles can impact inelastically.
This is affirmed by the poor film adhesion using 100\% ZnS, which is easily brushed off.   
Addition of ZnS into the diamond powder may provide a more inelastic impact, effectively raising $KE_b$.  
With increasing ZnS percentage less diamond is likely present in the aerosol, but the total amount of diamond adhering to the substrate is increased since there is more overall material deposited. 
 Furthermore, diamond even at higher percentages of ZnS is clearly interacting with the substrate and producing a rough surface which is likely continuing to facilitate the adhesion of both ZnS and diamond (see Figure \ref{mix-dep} (b) and (d)).  
Inherent in this process are the mechanical properties materials.
 The substrate plasticity, evident in the SEM images, may explain why film growth occurs on silicon and sapphire, but not ZnS.
 Local plasticity has been found to occur in both sapphire and silicon at pressures of 10s of GPa, which coincide with particle impact pressures during ADM\cite{akedo-room}. 
 ZnS, which is about 5 times softer than silicon may not provide adequate plasticity at these pressure to withstand the particle bombardment.  This motivates several avenues of study.  Two such studies might be on surface treatment of the ZnS substrate and decreasing the particle impact velocity.

The combined data from the SEM, profilomentry profiles, FTIR spectra, and XPS spectra suggest that there may be another mechanism involved in the deposition process than simply abrasion of the substrate. 
 SEM images show that the local surface structure across all mixtures is similar. 
Measurements of rms roughness confirm that only a slight increase in roughness occurs in going from 0\% -- 100\% (data not shown).  
XPS data show a clear Zn presence on the surface at all but the 0\% mixture and transmittance variance suggests a change in the bulk material. 
 A possible mechanism could be compression of the substrate surface due to high pressure bombardment of particles. 
 Reports of nano-indentation of less than 10 GPa onto silicon have shown a structural phase change which results in a 21\% volume collapse. 
Rapid compression and release could result in a meta-stable structural phase with a substantially smaller volume accounting for the apparent abrasion and existance of particles adhered to the region. 
 
There are several strategic avenues to pursue to address many of the questions still outstanding in this work.  One such study would be to develop a thin impact absorbing film coating on ZnS to increase $KE_b$. The energy absorbed by the film during impact may help decrease abrasion and micro-fracturing in the substrate.
Additionally, lower energy impact using smaller particles and lower velocities might improve substrate abrasion.
Another approach might be to incorporate the properties of ZnS and diamond together as a core/shell particle.
An advantage of this approach would be to lower the impact of each diamond particle and facilitate adhesion both with the substrate and other core/shell particles.
Abrasion and wear testing is an obvious neccesity in future work.
Comparing the as-deposited films with those sintered in an oxygen-free environment or by hot isostatic pressing may improve film density and abrasion resistance.

In conclusion, we utilized a newly constructed system to perform the aerosol deposition method of impacting diamond and mixtures of ZnS/diamond onto silicon, sapphire, and ZnS substrates to better understand the deposition mechanism with the goal of achieving deposition onto ZnS.
The results of this work suggest that film formation depends on particle velocity ($\Delta$P), substrate mechanical properties, and ZnS/diamond percentage.
We achieved a maximum film thickness of 0.9 (0.7) $\mu$m onto silicon (sapphire) with a 90\% ZnS mixture likely facilitated by the  local mechanical structure of the silicon and sapphire and sustained substrate deformation by the diamond.
All the targeted deposition areas showed very high transmittance in the 4 -- 7 $\mu$m wavelength range with a constant surface roughness.

Future work to employ a suface coating on ZnS, using smaller particles, decreasing the impact velocity, and synthesizing a core/shell particle may provide beneficial to improving abrasion and producing thicker films. A future film study will also include sintering and abrasion resistance testing. 

\section*{Acknowledgments}
SDJ gratefully acknowledges the support of the American Association for Engineering Education/ NRL Postdoctoral Fellowship program and  Ron Holm for his part in the design and implementation of the NRL ADM system.
Work at the U.S. Naval Research Laboratory is supported by the Office of Naval Research.

\bibliography{spie-report}

\end{document}